\documentclass[]{aa}

\usepackage{txfonts}
\usepackage{graphicx} 
\usepackage{natbib} 
\usepackage{longtable} 
\usepackage{amssymb}
\usepackage{lscape}
\usepackage{rotating}
\usepackage[english]{babel}
\bibpunct{(}{)}{;}{a}{}{,}

\begin{document}

\title{The white dwarf cooling sequence of 47~Tucanae}

\author{Enrique Garc\'\i a--Berro\inst{1,2}\and 
        Santiago Torres\inst{1,2}\and
        Leandro G. Althaus\inst{3,4}\and
        Marcelo M. Miller Bertolami\inst{4,5,6}}
        
\institute{Departament de F\'\i sica Aplicada, 
           Universitat Polit\`ecnica de Catalunya, 
           c/Esteve Terrades 5, 
           08860 Castelldefels, 
           Spain
           \and
           Institute for Space Studies of Catalonia, 
           c/Gran Capita 2--4, 
           Edif. Nexus 104, 
           08034 Barcelona, 
           Spain
           \and
           Facultad de Ciencias Astron\'omicas y Geof\'isicas, 
           Universidad Nacional de La Plata,
           Paseo del Bosque s/n, 
           1900 La Plata, 
           Argentina
           \and
           Instituto de Astrof\'isica de La Plata, UNLP-CONICET,
           Paseo del Bosque s/n, 
           1900 La Plata, 
           Argentina
           \and
           Max-Planck-Institut f\"ur Astrophysik, 
           Karl-Schwarzschild Strasse 1, 
           85748 Garching,
           Germany
           \and
           Post-doctoral Fellow of the Alexander von Humboldt 
           Foundation}

\date{\today}

\titlerunning{The white dwarf cooling sequence of 47~Tuc}
\authorrunning{Garc\'\i a--Berro et al.}

\offprints{E. Garc\'\i a--Berro}


\abstract 
          {47~Tucanae is one of the most interesting and well observed
            and theoretically studied  globular clusters.  This allows
            us to study the reliability  of our understanding of white
            dwarf cooling sequences, to  confront different methods to
            determine  its   age,  and   to  assess   other  important
            characteristics, like its star formation history.}
          {Here we present a population synthesis study of the cooling
            sequence   of  the   globular   cluster  47~Tucanae.    In
            particular,  we   study  the  distribution   of  effective
            temperatures,  the shape  of the  color-magnitude diagram,
            and the corresponding magnitude and color distributions}
          {We  do so  using  an up-to-date  population synthesis  code
            based  on Monte  Carlo techniques,  that incorporates  the
            most recent and reliable cooling sequences and an accurate
            modeling of the observational biases.}
          {We find a good agreement between our theoretical models and
            the  observed data.  Thus, our  study, rules  out previous
            claims that there  are still missing physics  in the white
            dwarf   cooling  models   at  moderately   high  effective
            temperatures. We also derive the  age of the cluster using
            the termination of the  cooling sequence, obtaining a good
            agreement   with   the   age  determinations   using   the
            main-sequence turn-off.   Finally, we  find that  the star
            formation history  of the cluster is  compatible with that
            obtained  using main  sequence  stars,  which predict  the
            existence of two distinct populations.}
          {We  conclude that  a correct  modeling of  the white  dwarf
            population of globular clusters,  used in combination with
            the  number  counts of  main  sequence  stars provides  an
            unique tool to model the properties of globular clusters.}


\maketitle


\section{Introduction}

White dwarfs are the most usual stellar evolutionary end-point, and as
such they convey important and valuable information about their parent
populations. Moreover, their structure and evolutionary properties are
well  understood --  see, for  instance \cite{Alt2010a}  for a  recent
review -- and their cooling times are, when controlled physical inputs
are    adopted,   as    reliable    as    main   sequence    lifetimes
\citep{Salaris2013}.    These   characteristics   have   allowed   the
determination of accurate ages using the termination of the degenerate
sequence for both open and  globular clusters.  This includes, to cite
a   few  examples,   the   old,  metal-rich   open  cluster   NGC~6791
\citep{Gar2010}  which  has two  distinct  termination  points of  the
cooling sequence  \citep{Bedin05, Bedin08a, Bedin08b}, the  young open
clusters  M~67  \citep{Bellini}  and  NGC~2158  \citep{2158},  or  the
globular  clusters  M4  \citep{M4} and  NGC~6397  \citep{Hansen_2013}.
However,  the  precise shape  of  the  cooling sequence  also  carries
important information  about the  individual characteristics  of these
clusters, and  moreover can  help in checking  the correctness  of the
theoretical  white  dwarf   evolutionary  sequences.   Recently,  some
concerns -- based  on the degenerate cooling sequence  of the globular
cluster  47~Tuc --  have  been  raised about  the  reliability of  the
available  cooling  sequences  \citep{Goldsbury_2012}.   47~Tuc  is  a
metal-rich globular cluster, being  its metallicity [Fe/H]$=-0.75$ or,
equivalently,  $Z\approx0.003$.  Thus,  there  exist accurate  cooling
ages and progenitor evolutionary  times of the appropriate metallicity
\citep{Renedo_2010}.  Hence, this cluster can be used as a testbed for
studying the  accuracy and  correctness of the  theory of  white dwarf
evolution.   Estimates  of  its  age   can  be  obtained  fitting  the
main-sequence turnoff,  yielding values ranging from  10~Gyr to 13~Gyr
-- see \cite{2010AJ....139..329T} for a careful discussion of the ages
obtained fitting  different sets  of isochrones  to the  main sequence
turn off.  Additionally, recent estimates based on the location of the
faint turn-down of the white dwarf luminosity function give a slightly
younger age  of $9.9\pm0.7\,$Gyr \citep{Hansen_2013}.  Here  we assess
the  reliability   of  the  cooling  sequences   using  the  available
observational  data.   As it  will  be  shown below,  the  theoretical
cooling sequences agree well with this  set of data. Having found that
the theoretical white dwarf cooling sequences agree with the empirical
one  we determine  the absolute  age of  47~Tuc using  three different
methods, and  also we investigate  if the recent  determinations using
number counts of main sequence stars  of the star formation history is
compatible with the  properties of the degenerate  cooling sequence of
this cluster.


\section{Observational data and numerical setup}

\subsection{Observational data}

The set  of data  employed in  the present paper  is that  obtained by
\cite{Kalirai12},    which     was    also    employed     later    by
\cite{Goldsbury_2012}  to  perform their  analysis.   \cite{Kalirai12}
collected the photometry for white  dwarfs in 47~Tuc, using 121 orbits
of the Hubble Space Telescope (HST). The exposures were taken with the
Advanced Camera for Surveys (ACS) and  the Wide Field Camera 3 (WFC3),
and comprise 13 adjacent fields.  A detailed and extensive description
of the observations  and of the data reduction procedure  can be found
in  \cite{Kalirai12}, and  we  refer  the reader  to  their paper  for
additional details.

\subsection{Numerical setup}

We use  an existing Monte  Carlo simulator which has  been extensively
described in  previous works \citep{MC1,MC2,MC3}.   Consequently, here
we will only summarize the ingredients which are most relevant for our
work.  Synthetic main  sequence stars are randomly  drawn according to
the initial  mass function  of \cite{Kroupa}.   The selected  range of
masses is  that necessary  to produce the  white dwarf  progenitors of
47~Tuc.   In  particular,  a  lower  limit of  $M  >  0.5\,  M_{\sun}$
guarantees that enough white dwarfs are  produced for a broad range of
cluster  ages.   In  our  reference  model we  adopt  an  age  $T_{\rm
c}=11.5$~Gyr, consistent with the main-sequence turn-off age of 47~Tuc
-- see \cite{Goldsbury_2012}  and references therein.  We  also employ
the star  formation rate of  \cite{Ventura_2014}, which consists  in a
first burst of star formation of duration $\Delta t=0.5$~Gyr, followed
by a  short period  of time  ($\sim 0.04$~Gyr)  during which  the star
formation activity ceases, and a  second short burst of star formation
which  lasts  for  $\sim  0.06$~Gyr.   The  fraction  of  white  dwarf
progenitors that are  formed during the first burst  of star formation
is 25\%, whereas the rest of  the synthetic stars (75\%) -- which have
a  helium enhancement  $\Delta Y\sim  0.03$ --  are formed  during the
second   one.    According    to   \cite{Ventura_2014}   the   initial
first-population in 47~Tuc  was about 7.5 times more  massive than the
cluster current total mass.  At present, only 20\% of the stars belong
to   the   population   with   primeval   abundances   \citep{Milone}.
Consequently, there  is a  small inconsistency in  the first-to-second
generation number  ratio employed  in our  calculations. Since  we are
trying to reproduce the  present-day first-to-second generation ratio,
this inconsistency might have consequences in our synthetic luminosity
function. To check this we conducted an additional set of calculations
varying the percentage  of stars with primeval abundances  by 5\%, and
we  found  that  the  differences in  the  corresponding  white  dwarf
luminosity functions were negligible.

Once we know which stars had time to evolve to white dwarfs we compute
their photometric  properties using the theoretical  cooling sequences
for  white dwarfs  with  hydrogen  atmospheres of  \cite{Renedo_2010}.
These  cooling  sequences  are  appropriate because  the  fraction  of
hydrogen-deficient  white  dwarfs  for   this  cluster  is  negligible
\citep{Woodley_2012}.   These  evolutionary   sequences  were  evolved
self-consistently  from  the  ZAMS,   through  the  giant  phase,  the
thermally  pulsing AGB  and mass-loss  phases, and  ultimately to  the
white dwarf  stage, and encompass  a wide  range of stellar  masses --
from  $M_{\rm  ZAMS}=0.85$  to   $5\,M_{\sun}$.   To  obtain  accurate
evolutionary  ages  for  the  metallicity  of  47~Tuc  ($Z=0.003$)  we
interpolate the cooling ages between the solar ($Z=0.01$) and subsolar
($Z=0.001$)   values   of   \cite{Renedo_2010}.    For   the   second,
helium-enhanced, population of  synthetic stars we computed  a new set
of  evolutionary sequences  which encompass  a broad  range of  helium
enhancements.  Finally, we also interpolate the white dwarf masses for
the   appropriate   metallicity   using  the   initial-to-final   mass
relationships of \cite{Renedo_2010}.

\begin{figure}[t]
   \resizebox{\hsize}{!}
   {\includegraphics[width=\columnwidth]{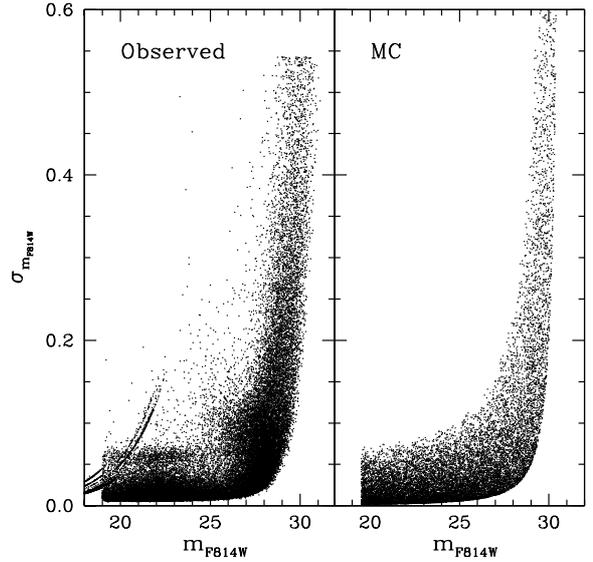}}
   \caption{Observed -- left panel --  and simulated -- right panel --
     distributions of photometric errors. See text for details.}
\label{f:errors}
\end{figure}

Photometric  errors are  assigned randomly  according to  the observed
distribution.   Specifically,  for  each  synthetic  white  dwarf  the
photometric errors  are drawn within a  hyperbolically increasing band
limited  by  $\sigma_{\rm l}=0.2\left(m_{\rm  F814W}-31.0\right)^{-2}$
and  $\sigma_{\rm   u}=1.7\left(m_{\rm  F814W}-31.0\right)^{-2}+0.06$,
which fits  well the  observations of  \cite{Kalirai12} for  the F814W
filter.  Specifically,  the photometric errors are  distributed within
this   band   according   to   the   expression   $\sigma=(\sigma_{\rm
u}-\sigma_{\rm  l})x^2$, where  $x\in(0,1)$ is  a random  number which
follows  an  uniform  distribution..   Thus,  the  photometric  errors
increase  linearly   between  the  previously   mentioned  boundaries.
Similar  expressions are  employed for  the rest  of the  filters. The
observed and  simulated photometric  errors of  a typical  Monte Carlo
realization are compared in  Fig.~\ref{f:errors} for the F814W filter.
As can  be seen  in this  figure, the  observed and  the theoretically
predicted  distribution  of  errors  display a  reasonable  degree  of
agreement.    In  particular,   the  observed   and  the   theoretical
distributions of photometric  errors have a relatively  broad range of
values for  F814W ranging from about  20~mag to 25~mag, for  which the
width of the  distribution remains almost flat. However,  the width of
the distribution  increases abruptly for  values of F814W  larger than
this last value.


\begin{figure}[t]
   \resizebox{\hsize}{!}
   {\includegraphics[width=\columnwidth]{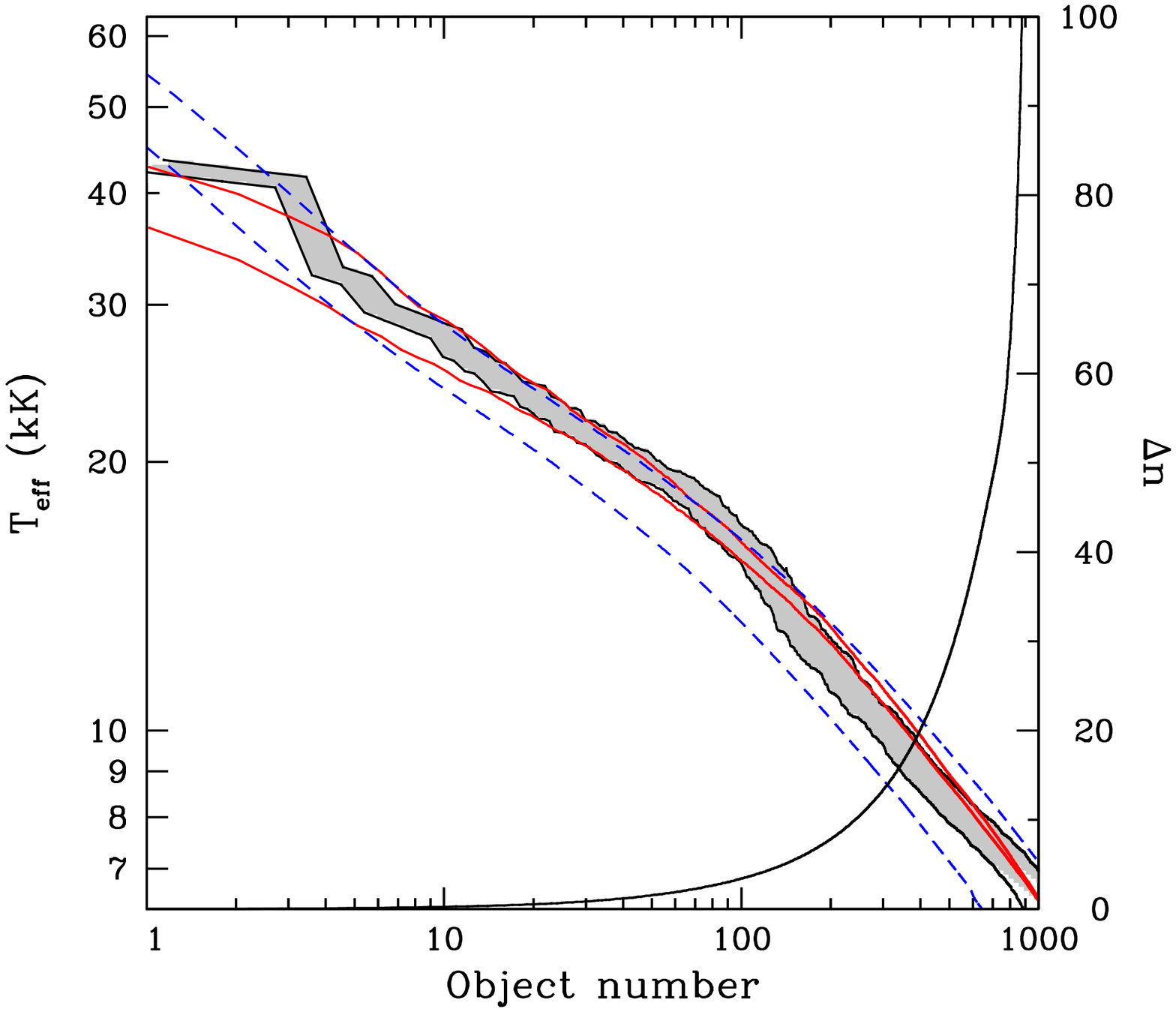}}
   \caption{Distribution of  effective temperatures  as a  function of
     the  white dwarf  object number  (left axis).   The observational
     values of \cite{Goldsbury_2012} are displayed using a grey shaded
     area, while the results of  our Monte Carlo simulations are shown
     using red lines.  The blue dashed lines show the cooling sequence
     for  $M_{\rm  WD}=0.53\,   M_{\sun}$  of  \cite{Renedo_2010}  for
     $Z=0.001$ under  different assumptions. The correction  factor of
     the  observed sample  of white  dwarfs is  also displayed  (right
     axis).  See the online edition of the journal for a color version
     of this plot.}
\label{f:Teff}
\end{figure}

\begin{figure}[t]
   \resizebox{\hsize}{!}
   {\includegraphics[width=\columnwidth]{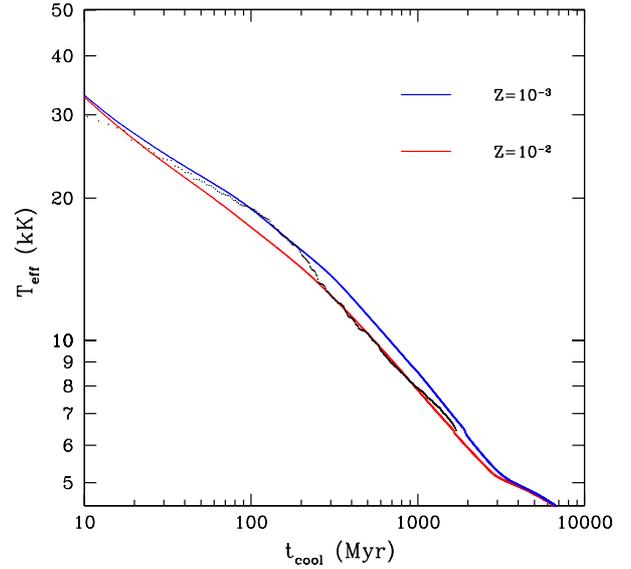}}
   \caption{Cooling sequences  of \cite{Renedo_2010}  for the  mass of
     the white dwarf corresponding to the main-sequence turn-off mass,
     and  for two  metallicities,  compared to  the empirical  cooling
     sequence of \cite{Goldsbury_2012}.}
\label{f:cooling}
\end{figure}

\begin{figure}[t]
   \resizebox{\hsize}{!}
   {\includegraphics[width=\columnwidth]{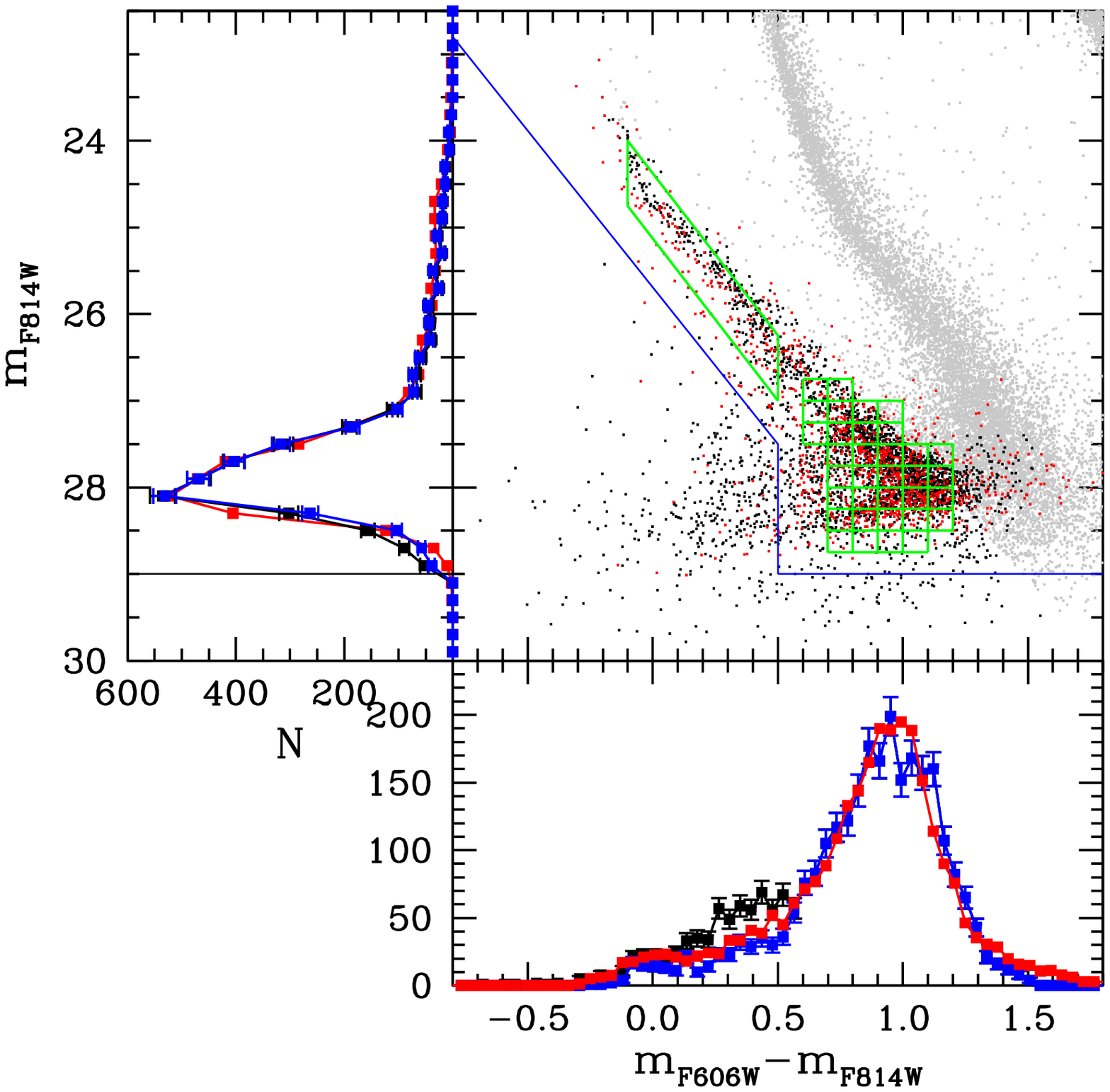}}
   \caption{White dwarf  luminosity function,  color-magnitude diagram
     and color  distribution of 47~Tuc.  Grey  dots represent observed
     main  sequence  stars,  black  dots correspond  to  white  dwarfs
     observed in the field, while red points denote the results of our
     synthetic  population   of  white  dwarfs.   The   green  squares
     represent  the  regions  of  the color-magnitude  for  which  the
     $\chi^2$ test was performed, while the blue thin lines correspond
     to the cuts adopted to  compute the distributions. The red curves
     correspond to  the simulated distributions obtained  when no cuts
     are  adopted,  the  blue  ones  are  the  observed  distributions
     computed using our  cuts, while the black lines  are the observed
     distributions when no  cuts are employed. See  the online edition
     of the journal  for a color version of this  figure, and the main
     text for additional details.}
\label{f:WDLF}
\end{figure}

\section{Results}

\subsection{The empirical cooling curve}

To start  with, we discuss  the distribution of white  dwarf effective
temperatures, and we compare it  with the observed distribution, which
is displayed in Fig.~\ref{f:Teff}.  \cite{Goldsbury_2012} measured the
effective temperatures  of a large  sample of white dwarfs  in 47~Tuc.
Afterwards they produced  a sorted list, from the hottest  star to the
coolest  one, to  experimentally  determine the  rate  at which  white
dwarfs are  cooling.  They  assumed a  constant white  dwarf formation
rate. Thus,  the position in  the sorted  list is proportional  to the
time    spent    on    the   cooling    sequence.     Their    sorted,
completeness-corrected distribution of effective temperatures is shown
Fig.~\ref{f:Teff} as a grey shaded  band, which includes the $1\sigma$
statistical  errors.   Also shown  for  illustrative  purposes is  the
completeness correction (solid black line).

The most salient feature of the observed distribution is the existence
of a pronounced break, which  occurs at effective temperatures $T_{\rm
eff}=20,000$~K.     This     feature    remained     unexplained    in
\cite{Goldsbury_2012} --  see their  Fig.~10 --  and prompted  them to
attribute  its origin  to some  missing piece  of physics  in all  the
existing  models  at  moderate   temperatures.   The  results  of  our
population  synthesis simulations  for  our reference  model are  also
shown, and  also include  the $1\sigma$ statistical  deviations (upper
and lower  red lines).  As can  be seen, our calculations  are in good
agreement with  the observed  data, without the  need of  invoking any
additional physical mechanism in  the cooling sequences, although they
do not  perfectly match the change  in slope of the  empirical cooling
curve.   In  the  following  we discuss  separately  several  possible
reasons that may  explain why our simulations are  in better agreement
with the observed distribution than those of \cite{Goldsbury_2012}.

The first reason that explains why  our model better fits the observed
list   is  that   we   used  updated   main-sequence  lifetimes,   the
initial-to-final  mass  relationship  of  \cite{Renedo_2010}  for  the
metallicity  of cluster,  and interpolated  cooling sequences  for the
precise  metallicity   of  47~Tuc,  while   \cite{Goldsbury_2012}  did
not. All  this results in a  different turn-off mass for  the cluster,
and consequently  in a different  white dwarf mass  at the top  of the
cooling sequence, hence  in different cooling rates.   To assess this,
in Fig.~\ref{f:cooling}  the sorted  list of  \cite{Goldsbury_2012} is
compared to the  theoretical cooling sequences of two  white dwarfs of
masses $M_{\rm WD}=0.525\, M_{\sun}$ and $0.520\, M_{\sun}$, the white
dwarf mass corresponding to the  main-sequence turn-off of 47~Tuc, for
two  metallicities  that  embrace  the  metallicity  of  the  cluster,
$Z=10^{-3}$ and $Z=10^{-2}$, respectively.   It is interesting to note
that the  $Z=10^{-3}$ sequence  provides a good  fit to  the empirical
cooling sequence of \cite{Goldsbury_2012} for temperatures hotter than
$T_{\rm eff}=20,000$~K, while  the $Z=10^{-2}$ sequence is  a good fit
for  temperatures colder  than this  value. Moreover,  the theoretical
cooling  sequences  bracket  the observed  distribution  of  effective
temperatures, and thus  it is not surprising that  our simulations fit
better the  observed distribution.   Finally, another  explanation for
the better fit of our simulations to the observed distribution is that
by construction  our simulations result  in a spread of  masses, while
\cite{Goldsbury_2012} adopted  a single cooling track  to compare with
their observational data.

Nevertheless, although our simulations  show an overall good agreement
with the observed distribution of  effective temperatures, they do not
fully reproduce the  observed break in the empirical  cooling curve at
moderately high effective  temperatures. We thus explore  which is the
origin  of this  discrepancy. Clearly,  metallicity cannot  be at  the
origin of  the break  because in  the region  of interest  the cooling
sequences of the white dwarfs  corresponding to the main turn-off mass
run almost  parallel.  Also,  we adopt the  star formation  history of
\cite{Ventura_2014}.  We repeated our  calculations employing a single
burst of  star formation and the  differences were found to  be minor.
Another possible origin for the break in the observed distribution may
be the sudden  increase photometric errors for  magnitudes larger than
25 --  see Fig.~\ref{f:errors}.  We thus  checked if the break  in the
distribution of observed errors is  responsible for the observed break
in sorted  list, and although  the inclusion of  realistic photometric
errors in the simulated population explains why our model is superior,
we found  that the observed  break cannot  be explained by  the sudden
increase in photometric  errors for stars with  magnitudes larger than
25~mag.   Thus, the  reason for  the observed  break in  the empirical
curve must be  related to either the way the  observed data is handled
or to an  unknown observational bias.  We  explore these possibilities
next.

In Fig.~\ref{f:Teff} we also show  the results of our simulations when
we adopt  the same  procedure used  by \cite{Goldsbury_2012}  -- upper
blue dashed  line.  This is equivalent  to adopt a single  white dwarf
cooling sequence of mass $M_{\rm  WD}=0.53\, M_{\sun}$ of a progenitor
with metallicity $Z=0.001$.  As can be seen, the first white dwarfs in
the  cooling curve  have  effective temperatures  larger than  $T_{\rm
eff}\sim    40,000$~K,    whereas    in    the    sorted    list    of
\cite{Goldsbury_2012}  none is  found.  If  white dwarfs  with $T_{\rm
eff}\ga 40,000$~K -- which  have very short evolutionary timescales,
and hence are difficult to detect  -- are removed from the theoretical
sorted list  the entire  distribution is shifted  towards the  left in
this diagram,  and we obtain  the lower blue  dashed line.  As  can be
seen this simple experiment helps  in solving the discrepancy found by
\cite{Goldsbury_2012}, although  does not totally remove  the reported
difference.   To better  assess this,  we employ  a Kolmogorov-Smirnov
test of  the cumulative  distributions of effective  temperatures.  We
first compute the statistic separation $D$, which measures the largest
separation between the cumulative  distribution of our simulations and
the observed data. The statistical  distances computed in this way are
0.0789 when  the Monte Carlo  simulation and the observed  sorted list
are compared,  0.2717 when the  model obtained using the  procedure of
\cite{Goldsbury_2012} is compared to the observed the sorted list, and
0.2694 when from this last model  the hottest white dwarfs are removed
from the sorted  list.  Using these statistical  distances, we compute
the probability of the three models being compatible with the observed
data. We  find that  the probability of  our Monte  Carlo distribution
being compatible with  the observational one is  $P\simeq 0.92$, while
this probability drops  to $P\simeq 0.73$ when  the procedure employed
by  \cite{Goldsbury_2012} is  adopted,  independently  of whether  the
hottest white  dwarfs are removed from  the sorted list or  not. Thus,
although  in a  strict statistical  sense none  of the  models can  be
totally  excluded to  a significant  level of  confidence --  say, for
instance,  5\% --  our population  synthesis model  presents a  better
agreement with  the observed  distribution of  effective temperatures,
and we  judge that  there is no  reason to invoke  a missing  piece of
physics at moderately high luminosities.

\subsection{The color-magnitude diagram}

Having assessed the reliability  of the theoretical cooling sequences,
we now discuss  the overall shape of the  color-magnitude diagram, and
the distributions  of magnitudes and  colors. All this  information is
displayed in Fig.~\ref{f:WDLF}. The central panel of this figure shows
the observed stars,  and the synthetic white dwarfs.  As  can be seen,
the distribution  of synthetic stars  perfectly overlaps with  that of
observed  ones,  except  at  very  low  luminosities,  for  which  the
contamination  with  galaxies is  very  likely.   For this  reason  we
introduced a  magnitude cut at  $m_{\rm F814W}=29$~mag, and we  do not
consider   these  objects   anymore   in   the  subsequent   analysis.
Additionally, since we do not have  proper motions of the white dwarfs
in 47~Tuc,  we introduced two more  cuts, which are also  displayed in
this figure.   These cuts  are also intend  to discard  all background
objects  which are  not cluster  members.   In particular,  we do  not
consider objects  to the left of  F606W$-$F814W=0.5~mag for magnitudes
between 29.0~mag  and 27.5~mag, and  objects to  the left of  the line
F814W=$3.62\left(\left({\rm    F606W}-    {\rm   F814W}    \right)-0.5
\right)+27.5$   for  brighter   magnitudes.    Since  the   background
contamination is also overimposed to  the white dwarf cooling sequence
we also estimated the number  of background contaminants still present
in the green boxes in Fig.~\ref{f:WDLF} . We did this in a statistical
way, by assuming that the density of contaminants within these regions
is similar to that close to the  exclusion line, and we found that the
percentage  of   contamination  of   each  of   the  green   boxes  in
Fig.~\ref{f:WDLF}  is   small,  $\sim  3\%$.   We   then  compute  the
theoretical white dwarf  luminosity function, and compare  it with the
observed luminosity functions when no  cuts are employed, and when the
color and magnitude cuts are  used.  Note that the differences between
the two  observed luminosity  functions are neglibible  for magnitudes
brighter than  that of the peak  at $\sim 28$~mag, and  very small for
fainter  magnitudes.   Moreover,  the  agreement  between  theory  and
observations  is  again  excellent.   We emphasize  that  should  some
physics was missing in the theoretical  cooling tracks we would not be
able to obtain  such a good agreement at  high luminosities.  Finally,
the bottom panel of this figure shows the color distributions.  Again,
the agreement is very good, except for the presence of a small bump in
the observed distribution at  F606W$-$F814W$\sim 0.3$~mag when no cuts
are used.  However,  when we discard the sources that  very likely are
not cluster white dwarfs the agreement is excellent.

\begin{table*}[t]
\begin{center}
\caption{$\chi^2$ test of the  luminosity function, color distribution
  and color-magnitude diagram, for different ages and durations of the
  first  burst of  star formation.  We  list the  normalized value  of
  $\chi^2$, that is the value of $\chi^2$ over its minimum value.}
\label{t:totalchic}
\centering
\begin{tabular}{crrrrrrrrrrrr}
\hline
\hline
\multicolumn{1}{c}{} &
\multicolumn{4}{c}{$\chi^2_{\rm F814W}/\chi^2_{\rm min}$} &
\multicolumn{4}{c}{$\chi^2_{\rm F606W-F814W}/\chi^2_{\rm min}$} &
\multicolumn{4}{c}{$\chi^2_{\rm N}/\chi^2_{\rm min}$}\\
\hline
\multicolumn{1}{c}{$T_{c}$ (Gyr)} &
\multicolumn{12}{c}{$\Delta t$ (Gyr)}\\
\cline{2-5}
\cline{6-9}
\cline{10-13}
  & 0.25 & 0.50 & 0.75 & 1.0 & 0.25 & 0.50 & 0.75 & 1.0 & 0.25 & 0.50 & 0.75 & 1.0\\
\cline{2-5} 
\cline{6-9}
\cline{10-13}
10.0 & 3.21 & 3.95 & 4.58 & 5.56 & 1.76 & 1.91 & 2.15 & 2.66 & 9.81 & 11.47 & 12.28 & 13.97  \\
10.5 &  1.96 &  2.44 &  2.97 &  3.57 &  1.29 & 1.44 & 1.63  & 1.81 & 5.93 & 7.67 & 9.22 & 10.46 \\
11.0 &  1.42 &  1.62 &  1.92 &  2.30 &  1.16 &  1.17 &  1.31  & 1.42 &  3.46 & 4.26 & 5.70 & 6.96 \\
11.5 &  1.25 &  1.27 &  1.33 &  1.42 &  1.00 &  1.00 &  1.03  & 1.10 & 1.49 & 2.08 & 2.95 & 3.81 \\
12.0 &  1.17 &  1.10 &  1.10 &  1.18 &  1.17 &  1.05 &  1.03 &  1.07 & 1.17 & 1.20 & 1.43 & 2.04  \\
12.5 &  1.53 &  1.35 &  1.13 &  1.00 &  1.59 &  1.39 &  1.14  & 1.11 & 1.49 & 1.31 & 1.00 & 1.19 \\
13.0 &  1.97 &  1.78 &  1.43 &  1.29 &  1.90 &  1.67 &  1.53 &  1.36 & 2.13 & 1.97 & 1.50 & 1.37 \\
\hline
\end{tabular}
\end{center}
\end{table*}

As  mentioned, the  white  dwarf cooling  sequence  of 47~Tuc  carries
interesting information about  its star formation history  and age. To
derive  this information  we use  the following  approach. We  compute
independent $\chi^2$  tests for  the magnitude  ($\chi^2_{\rm F814W}$)
and   color   ($\chi^2_{{\rm  F606W}-{\rm   F814W}}$)   distributions.
Additionally, we calculate  the number of white dwarfs  inside each of
the green boxes in the color-magnitude diagram of Fig.~\ref{f:WDLF} --
which are the same regions  of this diagram used by \cite{Hansen_2013}
to  compare  observations  and  simulations   --  and  we  perform  an
additional $\chi^2$ test, $\chi^2_{N}$.  We then investigate which are
the values of  the several parameters which define  the star formation
history of the  cluster -- that is,  its age, the duration  of the two
bursts,  and their  separation --  that  best fit  the observed  data,
independently.   That is,  we  seek  for the  parameters  of the  star
formation  history of  47~Tuc that  best  fit either  the white  dwarf
luminosity function, or the color  distribution or the number of stars
in each of the boxes  in Fig.~\ref{f:WDLF}.  Obviously, this procedure
results in  different values  of the parameters  that define  the star
formation history of 47~Tuc.

The  results of  our  analysis are  shown in  Table~\ref{t:totalchic},
where only the data for a reduced set of models in which we kept fixed
the  separation between  the  two  bursts of  star  formation and  the
duration of  the second burst, and  varied the age of  the cluster and
the duration of  the first burst, is listed.   We remark, nonetheless,
that we explored a significantly  larger range of parameters, and that
for the  sake of conciseness  we only show here  a few models.   We do
this because we find that the values of $\chi^2$ are less sensitive to
variations in  the rest  of parameters,  and thus  this set  of models
turns out to be quite representative.   As can be seen, when the white
dwarf luminosity function is employed to obtain the age of the cluster
and the  duration of  the burst  of star  formation the  $\chi^2$ test
favors an age $T_{\rm c}\simeq 12.5\pm 1.0$~Gyr and a duration $\Delta
t\simeq  1.0\pm  0.5$~Gyr. Instead,  when  the  color distribution  is
employed  we  obtain $T_{\rm  c}\simeq  11.5\pm  1.0$~Gyr and  $\Delta
t\simeq 0.6\pm 0.5$~Gyr, respectively, while  the model that best fits
the number of stars in each  bin of the color-magnitude diagram has an
age $T_{\rm c}\simeq  12.5\pm 0.5$~Gyr and a duration of  the burst of
star formation $\Delta  t\simeq 0.7\pm 0.5$~Gyr. These  results are in
accordance with those of  \cite{Ventura_2014}, agree with the absolute
age   determination  of   47~Tuc  using   the  eclipsing   binary  V69
\citep{2010AJ....139..329T},  and also  agreee each  other within  the
error bars.


\section{Conclusions}

In this  paper we have  assessed the  reliability and accuracy  of the
available cooling tracks using the white dwarf sequence of 47~Tuc.  We
have demonstrated that when the  correct set of evolutionary sequences
of the appropriate  metallicity are employed, and  a correct treatment
of  the photometric  errors,  and observational  biases  is done,  the
agreement  between   the  observed  and  simulated   distributions  of
effective temperatures, magnitudes and colors,  as well as the general
appearance of  the color-magnitude  diagram is excellent,  without the
need  of invoking  any missing  piece  of physics  at moderately  high
effective temperatures in the cooling  sequences.  While our models do
not  totally reproduce  the sudden  change of  slope in  the empirical
cooling sequence, it  is worth noting that such change  of slope takes
place in the  region where the completeness  correction factor becomes
relevant. Thus it might be well possible that this feature may be only
due to some unknown observational bias.

In a second phase, and given that  we found that there is no reason to
suspect the theoretical cooling sequences are incomplete, we also used
these distributions to study the age and star formation history of the
cluster  using   three  different   distributions:  the   white  dwarf
luminosity function, the color distribution,  and the number counts of
stars in  the color-magnitude  diagram. Using  these three  methods we
obtained that the age of the cluster is $T_{\rm c}\sim 12.0$~Gyr.  Our
results are compatible with the recent results of \cite{Ventura_2014},
who found  that star formation  in this cluster proceeded  through two
bursts, the first one of duration $\sim 0.4$~Gyr, while the second one
lasted  for $\sim  0.06$~Gyr, separated  by  a gap  of duration  $\sim
0.04$~Gyr.  We also found that  the relative strengths of these bursts
of  star formation  activity (25\%  and 75\%,  respectively), and  the
presence of  a helium-enhanced  population of white  dwarf progenitors
(born exclusively  during the second  burst) are also  compatible with
the characteristics of the white dwarf population.

Since our  analysis of the  cooling sequence of 47~Tuc  closely agrees
with  what  is obtained  studying  the  distribution of  main-sequence
stars, we conclude  that a combined strategy provides  a powerful tool
that can  be used to study  other star clusters, and  from this obtain
important  information  about  our  Galaxy.  In  these  sense,  it  is
important  to  realize that  \cite{Hansen_2013}  computed  the age  of
NGC~6397,  obtaining $\sim  12$~Gyr, significantly  longer than  their
computed  age  for 47~Tuc  ($\sim  10$~Gyr).   This prompted  them  to
suggest that  there is quantitative evidence  that metal-rich clusters
like  47~Tuc  formed later  than  the  metal-poor halo  clusters  like
NGC~6397. Our  study indicates  that 47~Tuc  is older  than previously
thought, and consequently, although this  may be true, more elaborated
studies are needed.


\begin{acknowledgements}
This work was partially supported by MCINN grant AYA2011--23102 by the
European  Union  FEDER  funds,  by AGENCIA  through  the  Programa  de
Modernizaci\'on   Tecnol\'ogica    BID   1728/OC-AR,   and    by   PIP
112-200801-00940  grant  from CONICET.   We  thank  R.  Goldsbury  and
B.M.S. Hansen  for providing us  with the observational data  shown in
Figs.~\ref{f:Teff} and \ref{f:WDLF}.
\end{acknowledgements}


\bibliographystyle{aa}
\bibliography{47_T}

\end{document}